\documentclass[a4paper,12pt]{article} 
\usepackage[margin=30mm]{geometry}

\usepackage[english]{babel}
\usepackage[pdftex]{graphicx} 
\usepackage[bf,small,nooneline]{caption}
\usepackage{citesupernumber} 
\usepackage{color}
\usepackage{multicol,lipsum}
\usepackage{amsmath}
\usepackage[document]{ragged2e}

\definecolor{darkgreen}{rgb}{0.2,0.6,0.4}

\author{K.~G.\ Kislyakova\thanks{
  Space Research Institute, Austrian Academy of Sciences, Schmiedlstrasse 6, A-8042 Graz, Austria. Email: kristina.kislyakova@oeaw.ac.at,  kristina.kislyakova@univie.ac.at}\thanks{
  University of Vienna, Department of Astrophysics,  T\"{u}rkenschanzstrasse 17, A-1180 Vienna, Austria.  
  }, 
L.\ Noack\thanks{Royal Observatory of Belgium, Department Reference Systems and Planetology, Avenue Circulaire 3, 1180 Uccle, Belgium.},
C.~P.\ Johnstone\footnotemark[2], 
V.~V.\ Zaitsev\thanks{Institute of Applied Physics, Russian Academy of Sciences, 46 Ul'yanov Street , 603950, Nizhny Novgorod, Russia.}, 
L.\ Fossati\footnotemark[1],\\
H.\ Lammer\footnotemark[1],
M.~L.\ Khodachenko\footnotemark[1],
P.\ Odert\footnotemark[1], and
M.\ G\"{u}del\footnotemark[2]}
\title{\textbf{\textsf{
Magma oceans and enhanced volcanism on TRAPPIST-1 planets due to induction heating
}}}
\date{\today}

\begin{document}
\maketitle

\textbf{
Low-mass M stars are plentiful in the Universe and often host small, rocky planets detectable with the current instrumentation. Recently, seven small planets have been discovered orbiting the ultracool dwarf TRAPPIST-1\cite{Gillon16,Gillon17}. We examine the role of electromagnetic induction heating of these planets, caused by the star's rotation and the planet's orbital motion. If the stellar rotation and magnetic dipole axes are inclined with respect to each other, induction heating can melt the upper mantle and enormously increase volcanic activity, sometimes producing a magma ocean below the planetary surface. We show that induction heating leads the three innermost planets, one of which is in the habitable zone, to either evolve towards a molten mantle planet, or to experience increased outgassing and volcanic activity, while the four outermost planets remain mostly unaffected.\\
} 


\section{The mechanism}

\begin{figure}
\includegraphics[width=0.7\columnwidth]{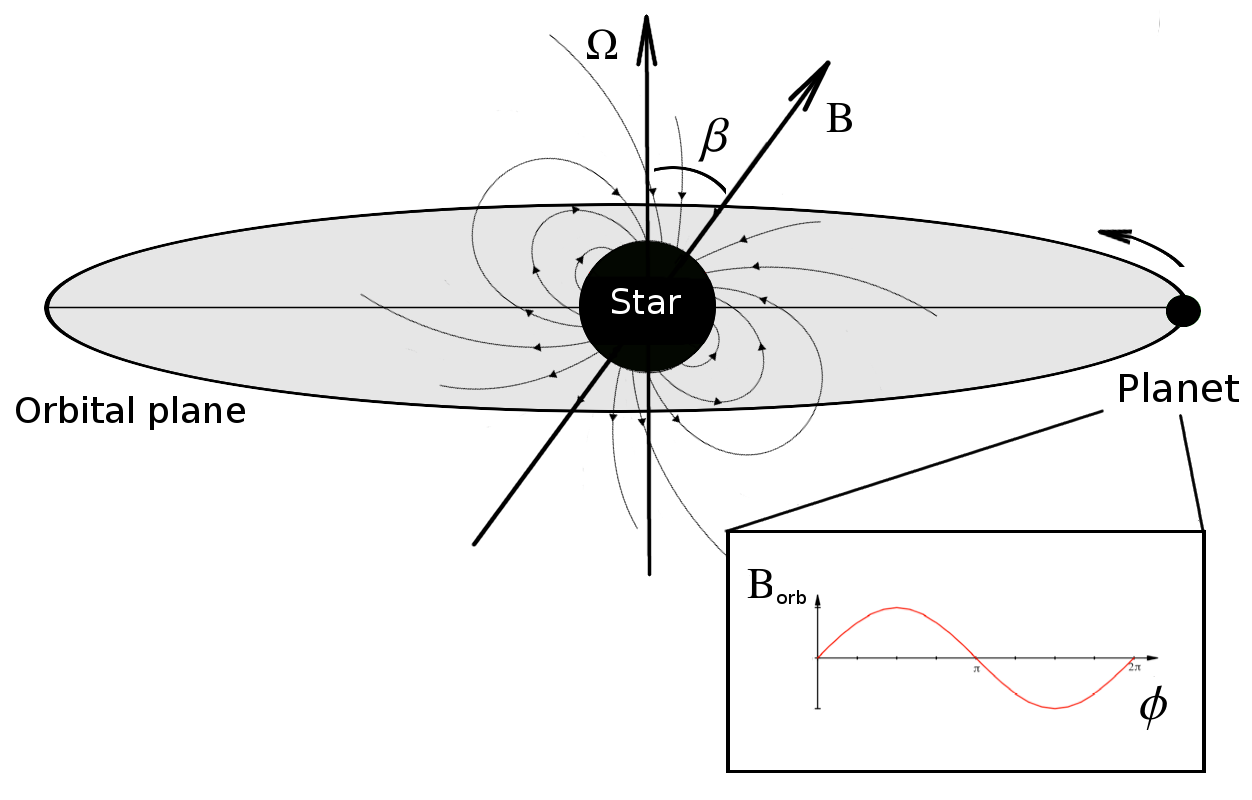}
\caption{ 
Sketch of the induction heating mechanism. If the stellar magnetic dipole is inclined with respect to the stellar rotation axis, the planet continuously experiences magnetic flux changes in its interior along its orbital motion. The change of the magnetic flux penetrating the planet generates eddy currents which dissipate and heat the planetary mantle. We assume that the magnetic field at the planetary orbit is changing harmonically with time and account for the dipole tilt by multiplying the field amplitude by $\sin \beta$, assuming $\beta = 60^{\circ}$. The stellar field lines at large radial distances are actually distorted outwards, and we account for the magnetic field to become radial at a certain distance from the star. In addition to the stellar magnetic field strength and planetary mantle parameters such as electrical conductivity, the interplay between stellar rotation and planetary orbital motion determines the heating. Further details can be found in Methods.
} \label{f_scetch}
\end{figure}

We propose induction heating of planetary interiors as an energy source in the planetary mantles. Induction heating arises when a changing magnetic field induces currents in a conducting medium which then dissipate to heat the body, mostly within an upper layer called the skin depth. Induction heating is widely used in various manufacturing processes to melt materials\cite{Rudnev03}. The same physical process can play a role in planetary interiors around strongly magnetized stars. Late M stars have been observed to have kG magnetic fields\cite{Morin10} that likely remain strong for billions of years, given the observed activity lifetimes of such stars\cite{West08}. Their habitable zones (HZ) are closer to the star, due to their low luminosity, which means that potentially habitable planets are exposed to much stronger fields than planets orbiting Sun-like G stars. The changing field generates currents in the planetary interior, which dissipate and heat the mantle. We study the effects of this mechanism on exoplanets. Some works\cite{Laine12} have considered the unipolar inductor heating of close-in exoplanets. In their model, a conducting body is moving inside a constant magnetic field, and the heating arises due to the motional electric field inside the body. The unipolar inductor model was originally applied to the Io-Jupiter system, where the two bodies are connected by a flux tube to close the electric circuit\cite{Goldreich69}. Here we consider a different model, where the change of the magnetic field arises due to the planetary orbital motion, stellar rotation, and stellar dipole tilt (Fig.~\ref{f_scetch}). This mechanism does not require a connecting flux tube, because in this case the alternating, and not direct, current is flowing inside the planet. We show that under certain conditions, induction heating can melt the planetary mantle, turning a planet into a body like Jupiter's moon Io with extreme volcanic activity and constant resurfacing. 

We calculate the energy dissipation caused by induction heating inside a planet in spherical geometry and consider a planet to be a sphere made up of concentric layers with uniform conductivity\cite{Parkinson83}. We solve the induction equation in every layer and find the magnetic field and current. Knowing the current and conductivity, we find the energy release within each layer. For details see Methods.


We apply our model to the planets orbiting the ultracool dwarf star TRAPPIST-1. This star has a mass of only 0.08~$M_{\odot}$, therefore, it seems plausible that it will decelerate very slowly and maintain a high level of magnetic activity for gigayears\cite{Irwin11}. The currently observed rotation period of TRAPPIST-1 is 1.4 days\cite{Gillon16}. We assume a dipole field with a strength of 600~G, which corresponds to the measured average magnetic field of TRAPPIST-1\cite{Reiners10}. The dipole component of the field may be (but is not necessarily) lower than the average value. We assume a fixed stellar magnetic field (both in strength and topology).  Table~\ref{t_1} summarizes the planetary parameters and calculated induction heating.

\section{Assumed conductivity and density profiles}
\label{sec_profiles}

\begin{figure}
\centering
\includegraphics[width=1.0\columnwidth]{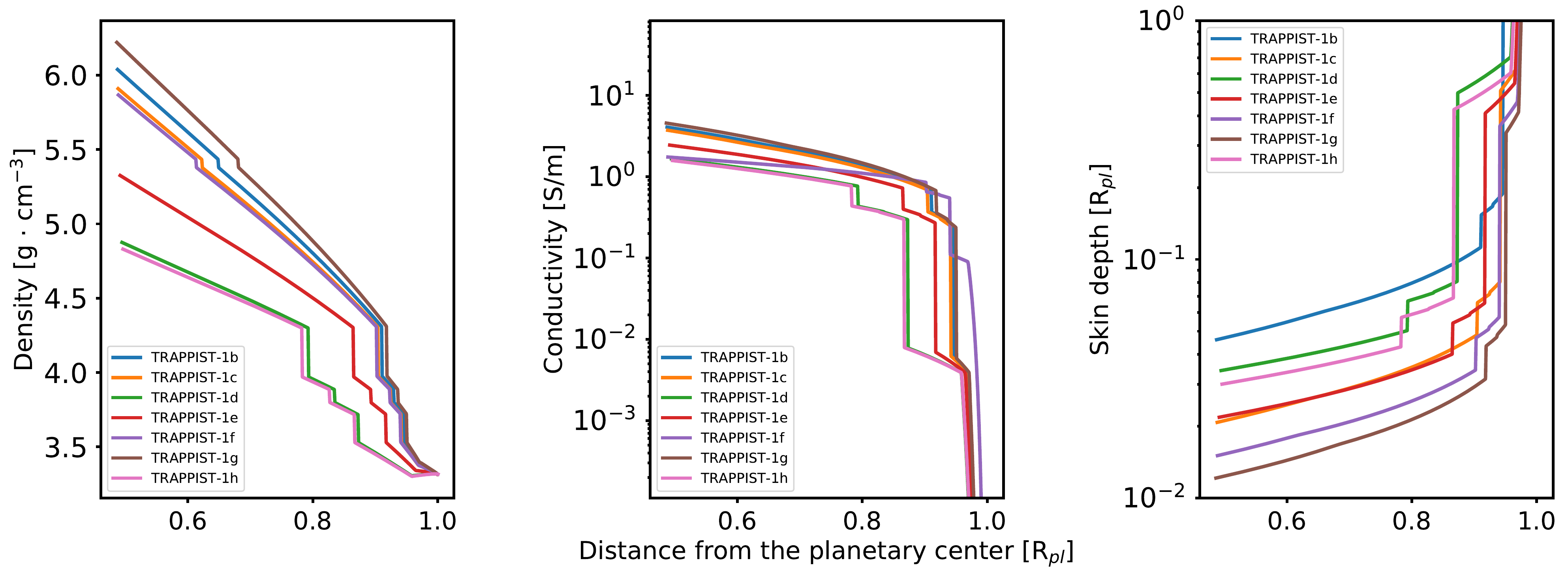}

\caption{Left panel: density profiles of all 7 TRAPPIST-1 planets. Middle panel: Conductivity profiles in S/m units for all TRAPPIST-1 planets, the distance from the planetary center is given in units of radii of each particular planet. 1 CGS unit = $1.11265 \times 10^{-10}$~Siemens/m. Right panel: radius dependent skin depth inside the TRAPPIST-1 planets calculated assuming the electrical conductivity shown in the middle panel. }  

\label{f_profiles}
\end{figure}

The conductivity profile is one of the important factors that influences the magnitude of the induction heating. Electrical conductivity of the medium determines the skin depth, $\delta$, which is the penetration depth of the electromagnetic field into the conducting medium, marking the level where the amplitude of the electromagnetic field decreases by a factor of $e$. Skin depth depends both on the conductivity of the medium and the period of the magnetic field change and is the smallest for a body with a high conductivity in a rapidly changing field. A constant magnetic field can fully penetrate the conducting body, therefore, $\delta$ equals infinity. In the examples we consider in this paper, we assume that the planets orbit in a vacuum, i.e., we disregard the skin effect in the ambient stellar wind plasma. In an approximation of a well conducting medium, namely, $4 \pi \sigma \gg \epsilon \omega$, which is true for the parameter range considered in the article, the skin depth is given by $\delta = c / \sqrt{2 \pi \sigma \mu \omega}$, where $c$ is the speed of light, $\sigma$ is the electrical conductivity of the medium, $\mu$ is the magnetic permeability, $\epsilon$ is the permittivity (hereafter we assume $\mu = \epsilon = 1$), and $\omega$ is the frequency of the field change. The skin depth is shown in the left panel of Fig.~\ref{f_profiles}.  It drastically decreases in the deeper layers, meaning that the magnetic field cannot penetrate the entire volume of the mantle. However, the skin depth in the upper layers of the mantle is comparable to or even exceeds the planetary radius. Both the depth and magnitude of the maximal heating depend on the skin depth. For small skin depth the maximum heating occurs close to the surface. On the contrary, if the skin depth is large, the field can penetrate deeper into the planetary mantle, so that the energy release occurs in deeper mantle layers. 

Since the masses of the TRAPPIST-1 planets are not yet precisely measured and little is known about the electrical conductivities of exoplanets, we assume an Earth-like composition and an Earth-like conductivity profile for a dry and iron-poor silicate mantle\cite{Xu00,Yoshino08,Yoshino13} (Fig.~\ref{f_profiles}, middle panel). We model planets in the stagnant lid regime, i.e., planets without plate tectonics or other catastrophic resurfacing mechanisms (like global mantle overturns), which is the dominant surface regime of rocky planets and moons in the Solar System. 

The electrical conductivity is calculated via
\begin{equation}
    \sigma = \sigma_0 \exp\left( - \frac{\Delta H}{k_B T} \right). 
    \label{eq_sigma}
\end{equation}
Here, $k_B$ is the Boltzman constant and $T$ the local temperature. 
The values for different phases of magnesium silicates are the following:
\begin{itemize}
	\item Olivine, $\sigma_0$=53703~S/m, $\Delta H$=2.31~eV \cite{Yoshino13},
	\item Wadsleyite, $\sigma_0$=399~S/m, $\Delta H$=1.49~eV \cite{Yoshino08},
	\item Ringwoodite, $\sigma_0$=838~S/m, $\Delta H$=1.36~eV \cite{Yoshino08},
	\item Perovskite, $\sigma_0$=11~S/m, $\Delta H$=0.62~eV \cite{Xu00},
	\item Magnesiowustite, $\sigma_0$=490~S/m, $\Delta H$=0.85~eV \cite{Xu00}.
\end{itemize}

The conductivity value of the perovskite-magnesiowustite phase is calculated via an arithmetic mean using the volume fractions as weights.

We have also calculated the density profiles with an interior structure model CHIC\cite{Noack16CHIC}, which were then used to calculate the local heating values in erg~s$^{-1}$~g$^{-1}$ (Fig.~\ref{f_profiles}, right panel). Phase transitions (olivine$\rightarrow$wadsleyite, wadsleyite$\rightarrow$ringwoodite, \\and ringwoodite$\rightarrow$perovskite mixed with magnesiowustite) are clearly visible in the density profile. We have considered the planets to have an Earth-like composition, but assumed the observed radii\cite{Gillon17} for the TRAPPIST-1 system. Given the uncertainties in the observed masses of the planets, assuming an Earth-like composition seems to be reasonable.

In this paper, we consider a ``dry'' conductivity profile. We also consider a terrestrial iron mass fraction of 35 wt-\%, which we assume accumulates entirely in the core. Iron-enrichment in the mantle would lead to an increase in the electrical conductivity of the mantle. Recently, a thick water-bearing level has been discovered in the Earth's mantle\cite{Pearson14}. The higher-pressure polymorphs of olivine, wadsleyite, and ringwoodite are able to host enough water to ensure an existence of a large water repository inside the Earth, hosting an amount of water far exceeding the one in the Earth's ocean. Such a water-rich layer in a planet's mantle can also be present in exoplanets, depending on their composition, accretion, and evolution. The presence of water in the planetary mantle influences the electrical properties of the upper mantle and leads to a different conductivity profile\cite{Yoshino13}, which we do not consider in the present study.

\section{Influence on interiors}
\begin{figure}
\centering
\includegraphics[width=1.0\columnwidth]{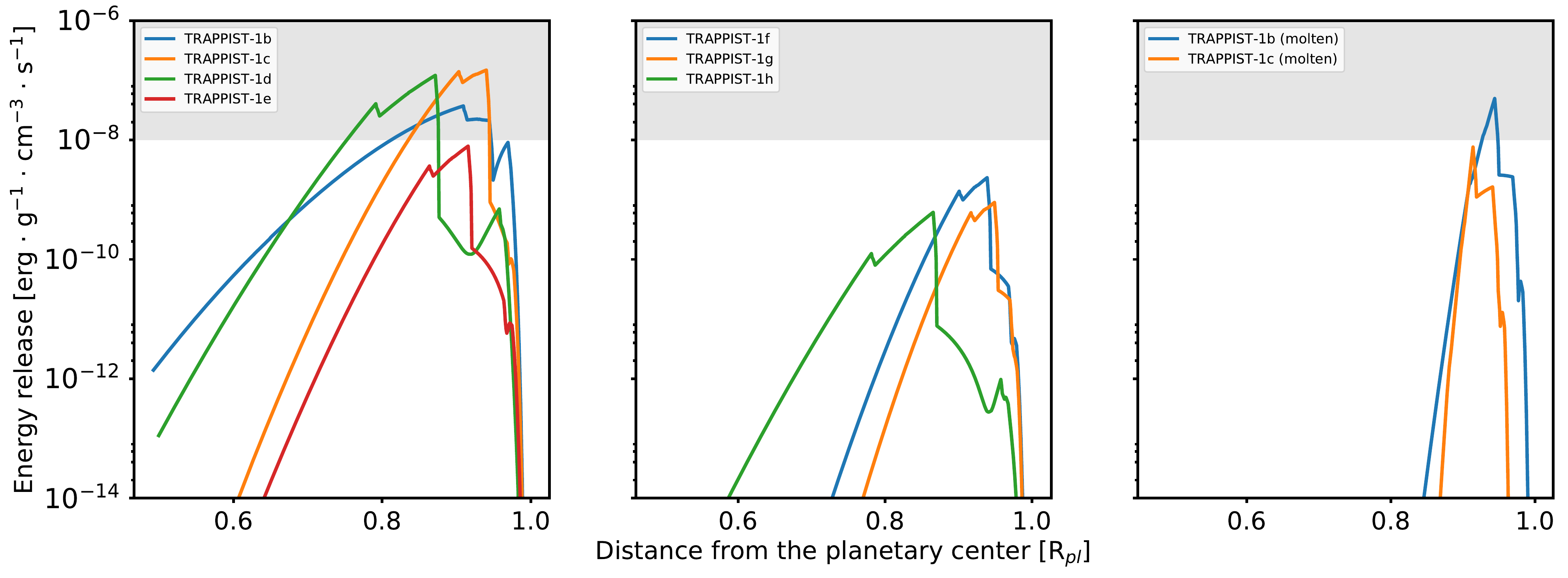}
\onecolumn
\caption{Induction heating inside the seven TRAPPIST-1 planets showing energy release rate inside the planets, normalized to the radius of each planet. The shaded area shows the approximate heating level necessary to keep the mantle molten.  Left panel: the heating rate inside planets TRAPPIST-1b, -c, -d, and -e. Middle panel: the same for planets TRAPPIST-1f, -g, and -h.
Right panel: illustration of the possible influence of an increased electrical conductivity of the molten mantle on the induction heating in planetary interiors. The electrical conductivity has the same profile as shown in the millde panel of Fig.~\ref{f_profiles} for TRAPPIST-1b and -c, but is multiplied by a factor of 100 to account for the increased melt fraction. The figure shows the energy release per mass unit. As one can see, the energy release decreases in comparison to the initial conductivity profile, but is still close to the energy level necessary to keep the mantle molten. The maximal heating shifts towards the planetary surface, because the magnetic field can not penetrate to the same depth inside a body with an increased conductivity.
}
\label{f_heating}
\end{figure}

We investigate if the induction heating of the mantle is sufficient to produce a magma ocean. Fig.~\ref{f_heating} shows the calculated induction heating rates (see Methods). One can see that the maximum heating occurs near the planetary surface at $\approx$0.9~R$_{\rm pl}$. The depth of the maximum heating depends both on the conductivity profile shape and on the period of the magnetic field variation. The latter is determined by the combination of the stellar rotation and planetary motion (see Methods). For this reason, TRAPPIST-1b shows has a lower energy release than TRAPPIST-1c, althouth the latter orbits further from the star. The orbital period of TRAPPIST-1b is close to the stellar rotational period, which leads to a lower heating rate.
Note that at the places where conductivity profiles shown in the middle panel of Fig.~\ref{f_profiles} have a jump, the heating profiles exhibit a jump too (at the border the heating drops to nearly zero). These heating drops are a pure computational effect and have therefore been removed. They do not affect overall heating values. 

\begin{figure}
\centering
\includegraphics[width=0.8\columnwidth]{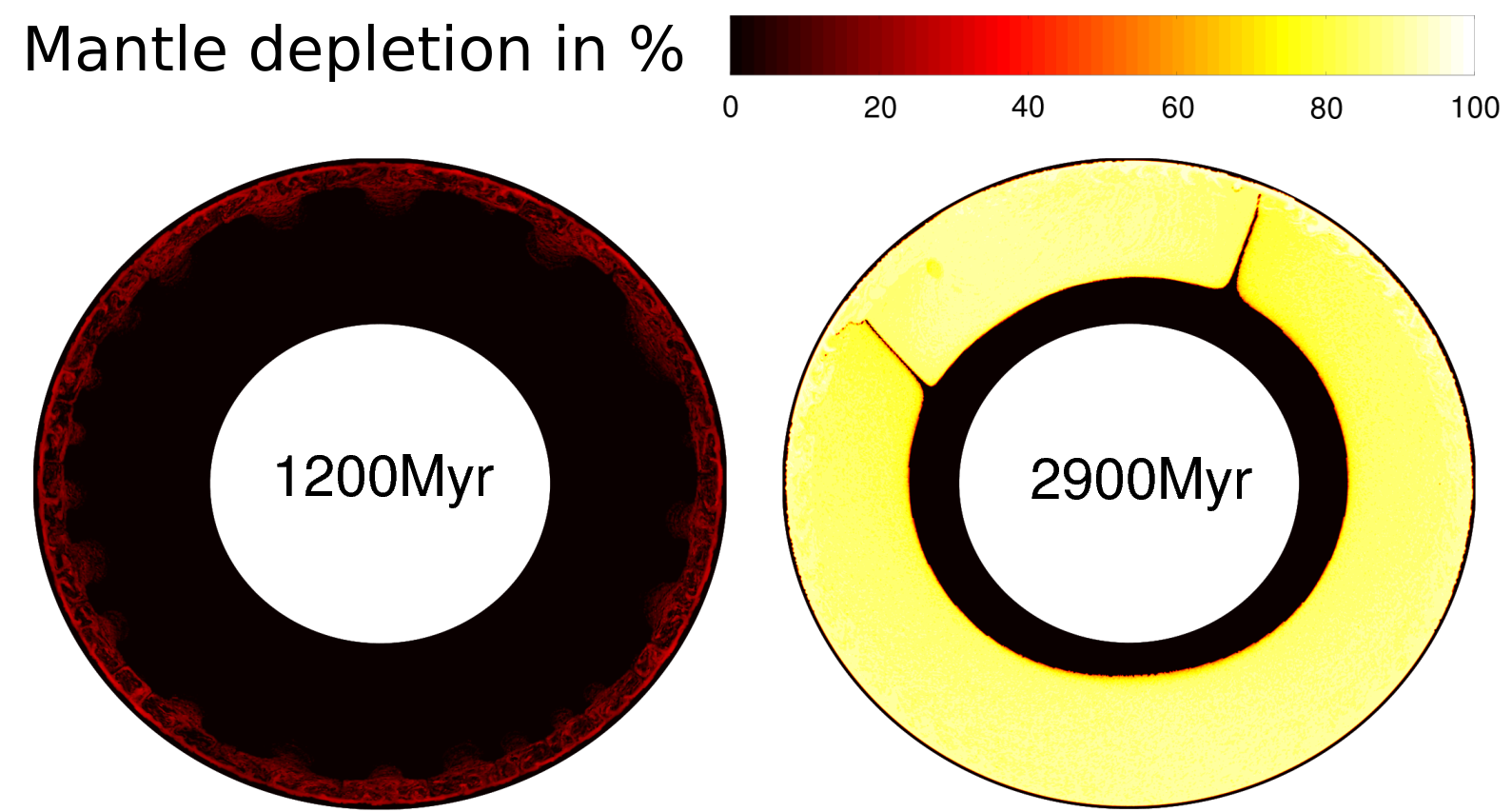}
\caption{Mantle depletion due to melting for TRAPPIST-1c (parameter case 1, see Methods for explanation). 
The three panels refer to system ages of 1.2, and 2.9~Gyr. The current age of the system is $\approx$500~Myr\cite{Gillon17}. Buoyant melting occurs only in a thin region below the lithosphere; depleted material is transported downwards by convection and new, primordial material (black) rises upwards. Complete melting occurs due to strong induction heating below the lithosphere (Fig.~\ref{f_heating}, left panel). 
Phase transitions in the mantle lead to inhomogeneous mantle mixing and plumes from the lower part of the mantle which are stable over geological time scales.}
\label{f_int}
\end{figure}

We use the finite volume code CHIC\cite{Noack16CHIC} to model the mantle convection and related magmatic events in a regional 2D spherical annulus geometry for a compressible mantle applying depth-dependent profiles for density, gravitational acceleration, thermal conductivity, electrical conductivity, thermal expansion coefficient, and specific heat capacity\cite{Noack16Melt}. We use in addition selected full 2D spherical annulus simulations (Fig.~\ref{f_int}) to confirm that the restriction to a regional domain does not influence the melting events.
Complete melting (i.e., where no solid particles flow in the magma) occurs if the temperature in the mantle locally exceeds the liquidus temperature, which is the temperature at which the entire rock is molten. We use the melting temperatures for peridotite\cite{Hirsch2000}, which resembles the primordial mantle material of Earth. 
For temperatures below the liquidus temperature, some minerals in the rock may melt (depending on the solidus, i.e. the minimal melting temperature), leading to partial melting of the mantle. Melt, if it is buoyant\cite{Ohtani95}, is assumed to penetrate through the lithosphere through dikes, and to lead to surface volcanism. Once melt solidifies at the surface, volatiles degas and lead to an enrichment of greenhouse gases in the atmosphere. Enhanced induction heating in the mantle, even if it does not lead to a magma ocean, still has an impact on the surface habitability by enhanced volcanic activity and outgassing of greenhouse gases. 
Strong magmatic events can gradually resurface the entire planet (similar to Io and possibly past Venus), which will have severe consequences for its habitability. 

\begin{figure}
\centering
\includegraphics[width=1.0\columnwidth]{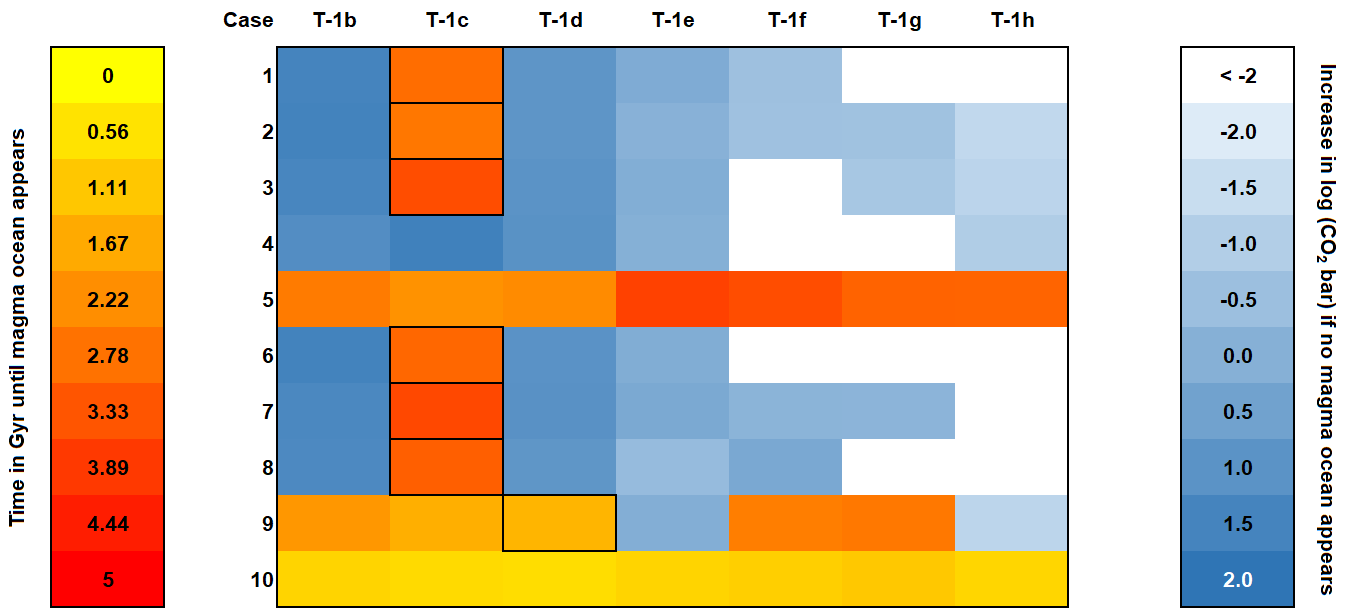}
\caption{Model results for 10 different example parameter cases (see Methods for explanation) 
investigating the influence of induction heating on the planets in the TRAPPIST-1 system assuming that they are comparable to Earth in terms of iron content, that their mantle consists of dry magnesium silicates, and assuming that they are stagnant-lid planets. We have varied parameters such as initial thermal boundary layer thickness, viscosity pre-factor, surface temperature and initial mantle temperature.
The yellow-red colour coding refers to the cases where the temperature in the upper mantle exceeds the liquidus temperature (i.e. a magma ocean appears). Thick boundaries indicate cases where a magma ocean forms when induction heating is considered, but is absent without induction heating.
The blue colour scale shows the logarithmic increase in total outgassed CO$_2$ in bar with respect to an evolution without induction heating after 5~Gyr of thermal evolution for cases where no magma ocean appears.}
\label{f_param}
\end{figure}

We trace the amount of volcanic outgassing produced during an evolution of 5~Gyr for the TRAPPIST-1 system. The left and middle panels of Fig.~\ref{f_heating} illustrate the energy release inside the TRAPPIST-1 planets. Taking into account this heating, we model mantle convection and melting processes for 10 different scenarios (cases in Fig.~\ref{f_param}, see Methods for description) to reflect unknown initial conditions and parameters. For some cases, local magma oceans form even in the absence of induction heating due to strong thermal effects and the absence of efficient mantle cooling by plate tectonics\cite{Reese98}. For some chosen parameters, a magma ocean appears for TRAPPIST-1c and -d due to induction heating (highlighted by thick black boundaries in Fig. \ref{f_param}). For cases with no magma oceans, the increase in outgassing (for CO$_2$)
assuming 20\% extrusive volcanism at the surface\cite{Crisp84} and 1000~ppm CO$_2$\cite{NiKe13} in the melt is plotted in Fig.~\ref{f_param} to highlight that induction heating leads to an increase in outgassing of greenhouse gases by up to several tens of bar, mainly for TRAPPIST-1b, -c, -d and -e. Fig.~\ref{f_int} illustrates the mantle melting (i.e. the amount of partial melt that occurred) inside TRAPPIST-1c at different times assuming constant induction heating. In Table~\ref{t_1}, we summarize the effect of induction heating compared to radiogenic heating (assuming Earth-like values), averaged over time. The maximal local heating effect (for TRAPPIST-1c) is 68.28\% of the local radiogenic heating, which explains why for this planet a magma ocean occurs for almost all parameter cases. The next-highest heating effect can be seen for TRAPPIST-1d (locally up to 55.75\%) and TRAPPIST-1b (up to 17.12\%). For the other four planets, induction heating is lower and has no influence on the thermal evolution of the planets.

\begin{table}[]
\begin{tabular}{lllllll}
\hline\hline
Planet       & Radius [R$_\oplus$] & Mass [M$_\oplus$] & Q$_m$ [erg/s] & Aver.		& Max. 		& B$_{\rm pl}$ [G]   \\
			 &						&					 &				  & $Q_m$/$Q_r$ & $Q_m$/$Q_r$  &   \\
\hline
b  & 1.086$\pm$0.035 & 1.315 (0.13-1.57) & $5.0 \times 10^{19}$ & 4.36\% & 17.12\%    & 1.37    \\
c  & 1.056$\pm$0.035 & 1.19 (0.77-1.99)  & $1.1 \times 10^{20}$ & 10.96\% & 68.28\%   & 0.73   \\
d  & 0.772$\pm$0.030 & 0.403 (0.14-0.68) & $2.5 \times 10^{19}$ & 6.91\% & 55.75\%    & 0.36   \\
e  & 0.918$\pm$0.039 & 0.73 (0.04-1.2)   & $2.7 \times 10^{18}$ & 0.42\% & 3.63\%     & 0.21   \\
f  & 1.045$\pm$0.038 & 1.15 (0.5-0.68)   & $1.1 \times 10^{18}$ & 0.12\% & 1.08\%     & 0.12   \\
g  & 1.127$\pm$0.041 & 1.5 (0.46-2.22)   & $5.3 \times 10^{17}$ & 0.04\% & 0.41\%     & 0.08   \\
h  & 0.755$\pm$0.034 & 0.375 (0-0.85)    & $\le 9.9 \times 10^{16}$ & $\le$0.03\% & $\le$0.28\%   & $\le$0.04   \\ \hline
\end{tabular}
\caption{Observed\cite{Gillon17} (with error bars) and modeled data for the TRAPPIST-1 system. For all planets we assume an iron weight fraction similar to Earth (35 wt-\%) and a dry, rocky mantle composed of magnesium silicates. We assume the observed radius and give the obtained mass for the assumed planet compositions together with the observed mass range in the 1$\sigma$ error interval. For TRAPPIST-1h, this is in agreement with a newly modelled mass constraint\cite{Luger17}. 
All model masses are within the observed mass range (i.e. all planets could be Earth-like, rocky planets) apart from TRAPPIST-1f. This planet may either have a true mass above the 1$\sigma$ measurement confidence level or could contain considerable amounts of water or ice leading to a lower planet mass. 
The last columns show total mantle-averaged induction heating $Q_m$, its ratio with respect to average radioactive heat sources $Q_r$ (assumed Earth-like) as well as the maximal ratio. Note that although the ratio $Q_m$/$Q_r$  averaged over the whole volume of the mantle is not very high, induction heating can still lead to high local values, because it concentrates in the upper mantle only (see Fig.~\ref{f_heating}).}
\label{t_1}
\end{table}

\section{Influence of the melt fraction on mantle conductivity}
\label{sec_melt}
Since the pressure and density profiles are mostly determined by the planetary gravity and composition, we assume that they do not change if the planet's mantle is molten. However, conductivity of the molten or partially molten rocks is known to increase both with pressure and temperature\cite{Yoshino06,Gaillard05,Maumus05} and melt fraction\cite{Yoshino10}. It has been shown\cite{Yoshino10} that electrical conductivity is increased by two or more orders of magnitude when the melt fraction is increased from 0.01 to 1. 

Although it is difficult to construct an exact electrical conductivity profile for an Earth-type planet with a fully molten mantle, we try to account for the influence of the high melt fraction and increased temperature for TRAPPIST-1b by multiplying the Earth-like conductivity profile in Fig.~\ref{f_profiles} by a factor of 100, based on experimental results\cite{Yoshino10}. One should note that this can be considered only as a coarse estimate.

The right panel of Fig.~\ref{f_heating} shows the energy release assuming this conductivity profile for the TRAPPIST-1b and -c planets, where a magma ocean could form. For simplicity, we assume here a global magma ocean extending from the surface to the core-mantle boundary. As one can see, in this case the skin effect is much stronger 
due to a higher conductivity. This leads to a shift of the heating maximum towards the planet's surface. When the mantle conductivity increases by two orders of magnitude, the total energy release decreases from $5.0 \times 10^{19}$ to $9.5 \times 10^{18}$~erg/s and from $1.1 \times 10^{20}$ to $1.1 \times 10^{18}$~erg/s for planets -b and -c, respectively. 

The shaded region in the Fig.~\ref{f_heating} denotes the approximate critical local value of the energy release necessary to melt the mantle. As one can see, near $\approx 0.95 R_{\rm pl}$ induction heating still provides values near the critical value or close to it. Therefore, we conclude that at least the upper planetary mantle will not always solidify once it is molten, although the lower mantle experiences no induction heating under these conditions. Depending on the density crossover and temperatures in the mantle, melt can sink to the lower mantle and form local melt ponds there\cite{Ohtani95,BeSch13}. The other possibility is the mantle solidifying and then being molten again.

\section{Discussion}
\label{sec_discussion}

We have investigated stagnant-lid planets, and with no internal dynamo. We have also not considered global mantle overturns or other catastrophic resurfacing mechanisms, which could efficiently remove the heat from the planet's mantle. Operating plate tectonics leads to subduction of cold lithosphere into the mantle and high heat flux at the surface and cools the mantle more efficiently than a stagnant lid\cite{BrMo15}, therefore, it might protect the planetary mantle from melting due to induction heating. On the other hand, induction heating may also inhibit the onset of plate tectonics, as it is believed that plate tectonics is favoured at lower internal temperatures\cite{StLoHa13,StBr14}. It is not clear if a larger size and mass of a planet would favour plate tectonics\cite{Val07,vanHeck11} or prevent it from operating\cite{StLoHa13,OnLe07,Noack14a}. An additional study is needed to determine the exact connection of induction heating and plate tectonics. Investigation of this issue is outside the scope of the present study. 

For Venus, other cooling scenarios have also been suggested, including lithosphere thickening which may have led to episodic subduction and mantle overturn events\cite{Sch97}, phase transition influencing whole-mantle convection by triggering catastrophic resurfacing\cite{Steinb93}, or a coupled atmosphere-interior feedback leading to ductile resurfacing events throughout Venus' history\cite{NoBrSp12}. In addition, a heat pipe mode as suggested for Io\cite{MoWe13} may lead to extraction of buoyant melt and enhance the cooling of the mantle by pressing the cold lithosphere down into the mantle. These mantle cooling processes may occur in the TRAPPIST-1 planets leading to a solidification of the magma ocean, but would have severe consequences for the planet's habitability, as they are related to resurfacing and strong volcanic activities. 

Induction heating has a certain influence on the dynamo in the Hermean core, when strong interplanetary magnetic fields and the solar wind environment under some conditions significantly amplify the dynamo, as is the case for Mercury\cite{Glassmeier07}. Mercury possesses a massive iron core where the dynamo is operating, so that the processes in the planetary mantle can be neglected. Earth-like planets which we consider here possess a much larger mantle in comparison to Mercury. The magnetic field decays in the mantle, being already insignificant at the core-mantle boundary. For this reason, we do not consider the influence of the induction on the planetary dynamo and focus on its heating effects only for non-magnetised planets. 

Our model does not take into account the evolution of the induction heating due to changing melt fractions in the planetary mantle. As we have shown above in Section~\ref{sec_melt}, increasing the melt fraction implies also an increase in the conductivity of the material, which would change also the skin depth and the heating distribution inside the body. To take it into account self-consistently is a complicated task, which requires a possible future study. 

Since strong external magnetic fields are necessary to make the induction heating substantial, this mechanism plays a negligible role for modern solar system objects, with the possible exception of Io, where the external magnetic field is provided by the tilted dipole field and the fast rotation of Jupiter\cite{Yanagisawa80}. The induction heating for Io is comparable to the energy release from radioactive decay\cite{Yanagisawa80}, but the spatial distributions of the two energy sources inside Io are different. For Io, the dominant energy source triggering the body's extreme volcanism is tidal heating\cite{Peale79}. 
The induction heating mechanism plays a much more important role for the planets orbiting late-type M dwarfs and, possibly, brown dwarfs, because it requires a close-in location of the planetary orbit for the magnetic heating to be strong enough.

Besides a strong magnetic field, a high frequency of the magnetic field variation is necessary for the induction heating to be substantial, i.e., a star has to be a fast rotator. Proxima~Centauri, which is the nearest star to the Earth and which has been proven to host an exoplanet in its habitable zone\cite{Anglada16}, presents an example of a slow rotating M dwarf with the observed rotational period of 83.5~days\cite{Anglada16}. Although its observed average magnetic field of 600~G\cite{Reiners08} is as strong as the one of TRAPPIST-1, the induction heating of Proxima~Centauri~b in its present evolutionary state is unsubstantial.

\section{Conclusions}
We have shown that induction heating plays an important role in the TRAPPIST-1 system. The two inner planets of the system likely evolve towards the molten mantle stage or enhanced occurence rate of magmatic events, and one planet in the habitable zone experiences increased volcanic activity and thus outgassing from the mantle. TRAPPIST-1e even shows a magma ocean for one parameter case, where no magma ocean would be found without induction heating.
Strong volcanic activity has been shown to have a severe negative influence on life and a planet's atmosphere\cite{Campbell92,Renne91}. For example, the eruptions from Siberian Traps could have led to acid rain, ozone layer depletion, and abrupt warming from greenhouse gas emissions, which altogether presented a severe stress on biota\cite{Black12,Black13}. Unlike the Earth, where volcanic eruptions of such scales have dire, but geologically short-lived, consequences due to the rareness of these events, for some TRAPPIST-1 planets, magmatism of similar or higher intensity could be constantly ongoing. On Io, the magma ocean under the surface triggers enormous volcanic activity, which is much stronger than that of the Earth\cite{McEwen98}. On the other hand, increased volcanic activity, if not extremely severe, may help the planet to keep its atmosphere or to replenish it if the planet loses it during the early most active phase of the stellar evolution. Therefore, the detrimental role of induction heating in the HZ of low-mass stars is not univocal and needs further investigations.

We conclude that the three innermost TRAPPIST-1 planets are affected by induction heating. TRAPPIST-1c and -d might possibly develop magma oceans in the course of their evolution, and in any event all three planets (-b, -c, and -d) should experience enhanced volcanic activity and outgassing. The four outermost planets, TRAPPIST-1e, -f, -g and -h are too far from the star for induction heating to be significant. We conclude that induction heating is an important factor for habitability in TRAPPIST-1 system, because it can drastically increase the amount of greenhouse gases, which, in their turn, influence the climate\cite{Alberti17,Wolf17}. We also conclude that induction heating should be in general taken into account to study the evolution of exoplanets orbiting low mass stars with strong magnetic fields.

\bibliographystyle{naturemag} 
\bibliography{references_NatAstron} 


\noindent\textbf{\textsf{Acknowledgements}}\\ 
We acknowledge the support by the FWF NFN project S116-N16 and the subprojects S11607-N16, S11604-N16, and S11606-N16. L.N. was funded by the Interuniversity Attraction Poles Programme initiated by the Belgian Science Policy Office through the Planet Topers alliance and by the Deutsche Forschungsgemeinschaft (SFB-TRR 170). This work results within the collaboration of the COST Action TD 1308. P.O. acknowledges FWF-project P27256-N27. V.Z. acknowledges RSF project 16-12-10528. The authors acknowledge the ISSI team ''The early evolution of the atmospheres of Earth, Venus, and Mars''. MLK also acknowledges the FWF projects I2939-N27, P25587-N27, P25640-N27, Leverhulme Trust grant IN-2014-016 and grant N 16-52-14006 of the Russian Fund of Basic Research.
\vspace{1em}

\noindent\textbf{\textsf{Author Contributions}}\\ 
K.K.\ calculated the induction heating. 
L.N.\ modeled the mantle effects. 
C.J.\ calculated stellar magnetic fields at a given orbital distance.
V.Z.\ helped to derive the equations. 
L.F.\ suggested that induction heating could be substantial.  
H.L.\ provided knowledge on the habitability of exoplanets.  
M.K.\ checked the influence of the ionosphere.  
P.O.\ helped writing the article.
M.G.\ contributed with expertise in exoplanetary research. 
All authors contributed to the text.
\vspace{1em}

\noindent\textbf{\textsf{Author Information}}\\ 
The authors declare no competing financial interests.
Correspondence and requests for materials should be addressed to K.K.\ 
(\texttt{kristina.kislyakova@univie.ac.at}).

\newpage
\onecolumn
\justify

\section*{Methods}

\textbf{Analytical estimate of induction heating.} Here we discuss the formulas used to calculate the induction heating.  We assume that a planet with the radius $R_{\rm pl}$ is embedded into an ambient magnetic field $\textbf{H} = \textbf{H}_{e} e^{-i \omega t}$, where $\textbf{H}_{e}$ is the dipole field strength at the planetary orbit (the external field), $\omega$ is the frequency of the magnetic field variation, and $i$ is the imaginary unit. We assume that the magnetic permeability $\mu = 1$, therefore, $\textbf{B} = \mu \textbf{H} = \textbf{H}$.  This assumption can be justified by several reasons. First, the Curie temperature, when the materials lose their magnetic properties (about 600$^\circ$ for olivine), is reached at a depth of only 20--90~km in our model depending on assumed planetary and mantle paratemeters, and at 40--60~km in the Earth mantle\cite{Ferre14}. Therefore, all mantle material below this depth is non-magnetic. Furthermore, althought the uppermost 20--90~km of the mantle could be magnetic\cite{Belley09,Nagata57}, the magnetic susceptibility, $\chi$, is of the order of $10^{-4}$--$ 10^{-3}$ for a typical density of the upper mantle (the latter value for a very high iron fraction\cite{Belley09}), thus giving $\mu = 1 + \chi \approx 1$. Therefore, we can assume $\mu=1$.

From the Maxwell equations, one can obtain the equation for the magnetic field:
\begin{equation}
\nabla^2 \textbf{B} = \frac{4 \pi \sigma \mu}{c^2} \frac{\partial \textbf{B}}{\partial t} + \frac{\epsilon \mu}{c^2} \frac{\partial^2 \textbf{B}}{\partial t^2}.
\end{equation}
If $4 \pi \sigma \gg \epsilon \omega$, the skin effect is important and the equation can be rewritten as
\begin{equation}
\nabla^2 \textbf{B} = \frac{4 \pi \sigma \mu}{c^2} \frac{\partial \textbf{B}}{\partial t} = k^2 \textbf{B}.
\label{e_mf1}
\end{equation}
If one assumes $B = B_e e^{-i \omega t}$ (a single sine wave of angular frequency $\omega$), then $k^2 = - i 4 \pi \omega \mu \sigma /c^2$, where $\sigma$ is the conductivity of the medium, and $c$ is the speed of light.

One can write a similar equation for the vector potential, $A$, which is defined as $\textbf{B} = {\rm rot} \textbf{A}$, namely, $\nabla^2 \textbf{A} = k^2 \textbf{A}$.


Inside the sphere, in a spherical coordinate system $[r, \theta, \phi]$, the magnetic field is given by\cite{Srivastava66,Parkinson83}:
\begin{equation*}
B_r = - (F(r)/r)  n(n+1) S_n^m
\end{equation*}
\begin{equation*}
B_\theta = - \frac{1}{r} \frac{d(rF(r))}{dr} \frac{\partial S_n^m}{\partial \theta} 
\end{equation*}
\begin{equation}
B_\phi = - (r \sin \theta)^{-1} \frac{d(rF(r))}{dr} \frac{\partial S_n^m}{\partial \phi},
\label{e_mfield}
\end{equation}
where the function $F(r)$ depends only on $r$ and where we have omitted the $e^{-i \omega t}$ term.


If the planetary mantle does not have a homogeneous conductivity, $k^2 \neq const$. Our approach is based on the formulas for a sphere made up of concentric shells each with uniform conducitivty, so that $k^2_j = const$ within a $j$-th layer\cite{Parkinson83,Srivastava66}. The sphere is divided into $m$ layers with the first layer being the planetary surface and the last $m$-th layer being the planetary core. One has to solve the induction equation within each layer and in the spherical central region. For a primary field outside the sphere described by a potential $U_e = B_e (r^n/R_{\rm pl}^{n-1}) S_n^m (\theta,\phi) e^{-i \omega t}$, where $B_e$ is the amplitude of the external field and $S_n^m(\theta,\phi)$ is a surface spherical harmonics, the magnetic field within the $j$-th layer can be expressed as\cite{Parkinson83}:

\begin{equation*}
B_r = (r k_j)^{-\frac{1}{2}}r^{-1}[C_j J_{n+\frac{1}{2}}(rk_j)+D_j J_{-n-\frac{1}{2}}(rk_j)]n(n+1)S_n^m
\end{equation*}
\begin{equation}
B_\theta = r^{-1}[C_jJ^*_{n+\frac{1}{2}}(rk_j)+D_jJ^*_{-n-\frac{1}{2}}(rk_j)]\partial S_n^m/\partial \theta
\label{e_mf}
\end{equation}
Here, $J$ is a Bessel function of a first kind, and $J^*_m(rk_j) = (d/dr)[(r/k_j)^{1/2} J_m(r k_j)]$. The derivations of the Bessel functions can be expressed as:
\begin{equation*}
J^*_{n+\frac{1}{2}}(rk_j) = (r k_j)^{1/2} J_{n-\frac{1}{2}}(rk_j) - n(rk_j)^{-1/2}J_{n+\frac{1}{2}}(rk_j)
\end{equation*}
\begin{equation}
J^*_{-n-\frac{1}{2}}(rk_j) = -(r k_j)^{1/2} J_{-n+\frac{1}{2}}(rk_j) - n(rk_j)^{-1/2}J_{-n-\frac{1}{2}}(rk_j)
\end{equation}
Therefore, the function $F(r)$ from Eq.~\ref{e_mfield} is given by $F(r_j) = C_j(r k_j)^{\frac{1}{2}}J_{n+\frac{1}{2}}(r k_j) + D_j(r k_j)^{\frac{1}{2}}J_{-n-\frac{1}{2}}(r k_j)$.

For a simple case of a homogeneous external field which we consider here, $n$ equals 1 and $S_n^m $ takes a simple form of $\cos \theta$, therefore, $U_e = B_e r S_1(\theta,\phi) e^{-i \omega t} = B_e r \cos \theta e^{-i \omega t}$. Substituting this into Eq.~\ref{e_mfield} one can immediately see that $B_\phi = 0$. In this case, the vector potential, which corresponds to the solution presented in Eq.~\ref{e_mf}, has only a $\phi$-component and is given by 
\begin{equation}
A_\phi = F(r)\sin \theta e^{-i \omega t}.
\label{e_A}
\end{equation}

The constants $C_j$ and $D_j$ are interconnected through following relations:

\begin{equation*}
C_j \alpha_j = C_{j+1} \beta_j + D_{j+1} \gamma_j,
\end{equation*}
\begin{equation}
D_j \alpha_j = C_{j+1} \delta_j + D_{j+1} \epsilon_j,
\label{e_cd}
\end{equation}
where 
\begin{equation*}
\alpha_j = J^*_{-n-\frac{1}{2}}(r_jk_j)J_{n+\frac{1}{2}}(r_jk_j) - J_{-n-\frac{1}{2}}(r_jk_j)J^*_{n+\frac{1}{2}}(r_jk_j),
\end{equation*}
\begin{equation*}
\beta_j = (k_j/k_{j+1})^\frac{1}{2}J_{n+\frac{1}{2}}(r_j k_{j+1})J^*_{-n-\frac{1}{2}}(r_j k_j)-J^*_{n+\frac{1}{2}}(r_j k_{j+1})J_{-n-\frac{1}{2}}(r_j k_j),
\end{equation*}
\begin{equation*}
\gamma_j = (k_j/k_{j+1})^\frac{1}{2}J_{-n-\frac{1}{2}}(k_j/k_{j+1})J^*_{-n-\frac{1}{2}}(k_j/k_{j})- J^*_{-n-\frac{1}{2}}(k_j/k_{j+1})J_{-n-\frac{1}{2}}(k_j/k_{j}),
\end{equation*}
\begin{equation*}
\delta_j = J^*_{n+\frac{1}{2}}(r_j k_{j+1})J_{n+\frac{1}{2}}(r_jk_j)-(k_j/k_{j+1})^\frac{1}{2}J_{n+\frac{1}{2}}(r_jk_{j+1})J^*_{n+\frac{1}{2}}(r_jk_j),
\end{equation*}
\begin{equation}
\epsilon_j = J^*_{-n-\frac{1}{2}}(r_jk_{j+1})J_{n+\frac{1}{2}}(r_jk_j)-(k_j/k_{j+1})^\frac{1}{2}J_{-n-\frac{1}{2}}(r_jk_{j+1})J^*_{n+\frac{1}{2}}(r_jk_j).
\end{equation}
If one introduces the relation $R_j = C_j/D_j$, one can obtain from Eq.~\ref{e_cd} a recursive formula
\begin{equation}
R_j = \frac{R_{j+1}\beta_j + \gamma_j}{R_{j+1}\delta_j + \epsilon_j}.
\end{equation}
In the center of the sphere, we can assume $k = k_m$, which in our case is the conductivity at the core-mantle boundary. To avoid singularity in the sphere's center, one has to assume $D_m = 0$, from which one gets the inital value for the recursive formula $R_{m-1} = \beta_{m-1}/\delta_{m-1}$. This gives the first boundary condition.

At the surface of the sphere one can obtain the relation between the amplitudes of the intrinsic and extermal field:
\begin{equation}
B_i/B_e = \frac{R_1 \beta_0 + \gamma_o}{R_1 \delta_0 + \epsilon_0},
\end{equation}
where
\begin{equation*}
\beta_0 = n[(ak_1)^{\frac{1}{2}}J^*_{n+\frac{1}{2}}(ak_1)-(n+1)J_{n+\frac{1}{2}}(ak_1)],
\end{equation*}
\begin{equation*}
\gamma_0 = n[(ak_1)^{\frac{1}{2}}J^*_{-n-\frac{1}{2}}(ak_1)-(n+1)J_{-n-\frac{1}{2}}(ak_1)],
\end{equation*}
\begin{equation*}
\delta_0 = (n+1)[(ak_1)^{\frac{1}{2}}J^*_{n+\frac{1}{2}}(ak_1)+nJ_{n+\frac{1}{2}}(ak_1)],
\end{equation*}
\begin{equation}
\epsilon_0 = (n+1)[(ak_1)^{\frac{1}{2}}J^*_{-n-\frac{1}{2}}(ak_1)+nJ_{-n-\frac{1}{2}}(ak_1)].
\end{equation}

At the surface of the sphere, where $r=R_{\rm pl}$, the magnetic field components equal:
\begin{equation*}
B_r = - [B_e n - (n+1) B_i] S_n^m = - [B_e - 2 B_i] \cos \theta
\end{equation*}
\begin{equation*}
B_\theta = -(B_e + B_i) \partial S_n^m/\partial \theta = (B_e + B_i) \sin \theta
\end{equation*}
\begin{equation*}
B_\phi = (B_e + B_i)(\sin \theta)^{-1}  \partial S_n^m/\partial \phi = 0.
\end{equation*}

For more details, we encourage the reader familiarize with the full derivation in the literature\cite{Parkinson83}.

To calculate the heating, we need to find the current flowing within each layer. 
The current can be found as ${\rm rot} \textbf{B} = \frac{4\pi}{c} \textbf{j}$ (NB: the layer number is also denoted by $j$, however, it is an index). On the other hand, ${\rm rot} \textbf{B} = \nabla^2 \textbf{A} = k^2 \textbf{A}$, where \textbf{A} is given by Eq.~\ref{e_A}. If we substitute $k^2$ in this equation, it gives
\begin{equation}
j_{\phi j} = - \frac{i \omega \mu \sigma_j}{c} F(r_j) \sin \theta e^{-i \omega t}
\end{equation}
as the only component of the current flowing within the $j$-th layer. The energy release in the layer can be calculated as

\begin{equation}
	Q_{j} = \frac{1}{2 \sigma_j} \int |j_{\phi j}|^2 dV
	\label{e_energy}
\end{equation}

In our calculations, we assume magnetic field change frequency:
\begin{equation}
	\omega = |\omega_{\rm st} - \omega_{\rm pl}|.
	\label{e_rot}
\end{equation}
Here, $\omega_{\rm st}$ is the frequency of the stellar rotation, and $\omega_{\rm pl}$ is the frequency of the planetary orbital motion. Both of them determine different skin depths shown in the right panel of Fig.~\ref{f_profiles}. We only consider prograde orbits in this paper. Although retrograde orbits are also possible, the majority of the planets are observed to have prograde orbits\cite{Winn15}. To account for a retrograde orbit, one should simply change Eq.~\ref{e_rot} to $|\omega_{\rm st} + \omega_{\rm pl}|$.

We assume that the planet is on a circular orbit around the star, so that the frequency of the orbital motion is determined by the stellar mass, $M_{\rm st}$, and the orbital distance, $R_{\rm orb}$. The speed of the orbital motion at a given distance is $v_{\rm orb} = \sqrt{G M_{\rm st}/R_{\rm orb}}$, where $G$ is the gravitational constant. This gives $\omega_{\rm pl} = \sqrt{G M_{\rm st} / R_{\rm orb}^3}$.

The orbital periods of habitable zone planets are longer for higher mass stars. For active rapidly rotating stars, the orbital periods are longer than the stellar rotation periods and lead to a small correction of $\omega$. In our article, we assume that TRAPPIST-1 never decelerates, which leads to constant induction heating inside the planets, if one also disregards the evolution of the stellar magnetic field. The influence of the evolution of the stellar rotation on the induction heating will be considered in a follow-up article.

In this paper, we have always assumed that the angle between the stellar rotation axis and the dipole axis, $\beta$, is 60$^{\circ}$. The value of $\beta$ influences the amplitude of the heating. Although the exact value of $\beta$ for TRAPPIST-1 is not available, observations of other low mass M dwarfs indicate that all angles are possible\cite{Lang12} and are probably variable in time.

\textbf{Estimating the magnetic field variations at a given orbital distance.} For our calculations, we are interested in the change of the magnetic field over the orbit of the planet. 
To do this, we use the Potential-Field Source-Surface model (PFSS), which is commonly used to model the magnetic fields of the Sun\cite{Altschuler69,Mackay02} and other stars\cite{Jardine02,Gregory11}.
In this model, the magnetic field is assumed to be potential (current-free) within the star's corona, which extends from the stellar surface to the `source surface' at $R_{ss}$, and then radial further from the star.
The assumption of a radial field at and beyond the source surface approximates the effects of the stellar wind on the magnetic field structure; it is significant because when \mbox{$r < R_{ss}$}, the field strength decreases with $r^{-3}$ for a dipole, and when \mbox{$r > R_{ss}$}, the field strength decreases with $r^{-2}$.

Specifically, we need to know \mbox{$\Delta B = B_\mathrm{max} - B_\mathrm{min}$}, where $B_\mathrm{max}$ and $B_\mathrm{min}$ are the maximum and minimum radial field strengths over the orbit. 
Since we are assuming a simple tilted dipole field geometry, \mbox{$B_\mathrm{max} = - B_\mathrm{min}$} and therefore \mbox{$\Delta B = 2 B_\mathrm{max}$}.
In the PFSS model, the magnetic field is described as the superposition of spherical harmonic components, each with degree $l$ and azimuthal order $m$ (see Eqs.~2.101--2.105 here\cite{JohnstoneThesis}).
The strength of each component is given by $B_{lm}$, with the dipole component corresponding to the three $l=1$ components (i.e. with $m=-1,0,1$). 
Since we are considering only planets orbiting in the equatorial plane (\mbox{$\theta=\pi/2$}), the $m=0$ component does not contribute, and we can ignore the $m=-1$ component by simply doubling the strength of the $m=1$ component (see section~2.3.3 in Johnstone\cite{JohnstoneThesis}).
In addition, we define our coordinate system so that the magnetic pole is at longitude \mbox{$\phi=0$}, meaning that the imaginary part of $B_{11}$ vanishes. 
The value of $B_{11}$ here is therefore the strength of the component of the dipole that is tilted 90$^\circ$ to the rotation axis. 
The magnetic field at the source surface as a function of $\phi$ is then given by \mbox{$B(R_{ss},\phi) = - 2 B_{11} f_1(R_{ss}) \cos\phi$}, where $f_1(R_{ss})$ is given by Eq.~2.104\cite{JohnstoneThesis}.
Since the maximum of $B(R_{ss},\phi)$ is where \mbox{$\cos\phi=1$}, we get
\begin{equation}
    \Delta B = 4 |B_{11}| \left[ \frac{3 (R_{ss}/R_\star)^{-3}}{(R_{ss}/R_\star)^{-3} + 2} \right] \left( \frac{r}{R_{ss}} \right)^{-2},
    \label{e_deltaB}
\end{equation}
where the $r^{-2}$ term scales the field strengths from the source surface to the planet's orbital radius, and $R_\star$ is the stellar radius.
The value of $B_{11}$ can be easily derived from the dipole field strength and tilt angle. 
Based on the example of the Sun, we assume \mbox{$R_{ss}=2.5 R_\star$}. The values of the magnetic field at the orbital distances of TRAPPIST-1 planets can be found in Table~\ref{t_1}.

\textbf{Influence of induction heating on planetary interiors.} We use fixed surface temperatures between 300 and 600 K (which results in a solid state of the rocks at the surface of these planets), and consider no temperature jump at the core-mantle boundary (CMB). A magma ocean or partial melting below the lithosphere leads to strong volcanic events at the surface and to strong outgassing of greenhouse gases.

To cover a broad range of possible initial states of the planet, we simulate ten different parameter cases based on a reference configuration, where we assume initial thermal boundary layers at the top and bottom of the mantle of 100~km, a dry reference viscosity\cite{KaWu93}, a fixed surface temperature of 300 K, and an initial upper mantle temperature of 1600~K at the bottom of the initial upper thermal boundary layer. he cases vary these parameters with assuming either 50 (case 2) or 200 km (3) thick initial thermal boundary layers, viscosity reduced (4) or increased (5) by one order of magnitude, surface temperatures of 400 K (6), 500 K (7) or 600 K (8). The last two cases use higher initial upper mantle temperatures of 1800 K (9) or 2000 K (10). 


The last two cases use higher initial upper mantle temperatures of 1800~K (9) and 2000~K (10). In both cases, the temperature lies initially above the solidus melting temperature and is cut down to the solidus profile to avoid initial melting. Since we neglect plate tectonics and heat piping in this study, the cases with higher initial mantle temperatures heat up efficiently even without any induction heating source, and the liquidus temperature is reached after an average of 2.1 and 0.9~Gyr, respectively (for TRAPPIST-1d, -1e and -1h, though, no magma ocean appears for case 9 without induction heating, since these planets are smaller than the other TRAPPIST planets and cool more efficiently). Similarly, an increased viscosity (Case 5) leads to less efficient heat transport in the mantle and a magma ocean is obtained after 2.9~Gyr on average.\\

\textbf{Possible protection by the ionosphere.} Even non-magnetised planets, such as Mars and Venus, possess ionospheres\cite{Lammer13,Terada09}. In the ionosphere, neutral atmospheric particles are ionised by stellar radiation and winds, providing a conducting medium. One should keep in mind that the  ionosphere is composed of plasma with varying ionization degree. In such a medium, the electrical conductivity is highly anisotropic so that field-aligned, Pedersen, and Hall conductivities can differ by many orders of magnitude. For the induction heating, the Pedersen component determines if the magnetic field change can reach the surface, because it is a conductivity component perpendicular to the magnetic field and along the tangential magnetic field component that plays a role in the reflection of an incident wave. In the Earth's ionosphere, the Pedersen conductivity is of the order of $10^{-6}$~S/m or $\approx 10^{4}$ in CGS units\cite{AuroraBook}. This value corresponds to a skin depth of $\approx$24~R$_\oplus$, which exceeds the size of the ionosphere by many orders of magnitude. Therefore, we conclude that the ionosphere cannot protect the planet from induction heating.

\section*{Data Availability}

The datasets generated during during the current study are available from the corresponding author on request.

\end{document}